\documentstyle[12pt]{article}
\def\journal{\topmargin .3in	\oddsidemargin .5in
	\headheight 0pt	\headsep 0pt
	\textwidth 5.625in % 1.2 preprint size  %6.5in
	\textheight 8.25in % 1.2 preprint size 9in
	\marginparwidth 1.5in
	\parindent 2em
	\parskip .5ex plus .1ex		\jot = 1.5ex}
\journal

\catcode`\@=11
\def\marginnote#1{}
\newcount\hour
\newcount\minute
\newtoks\amorpm
\hour=\time\divide\hour by60
\minute=\time{\multiply\hour by60 \global\advance\minute by-\hour}
\edef\standardtime{{\ifnum\hour<12 \global\amorpm={am}%
	\else\global\amorpm={pm}\advance\hour by-12 \fi
	\ifnum\hour=0 \hour=12 \fi
	\number\hour:\ifnum\minute<10 0\fi\number\minute\the\amorpm}}
\edef\militarytime{\number\hour:\ifnum\minute<10 0\fi\number\minute}
\def\draftlabel#1{{\@bsphack\if@filesw {\let\thepage\relax
   \xdef\@gtempa{\write\@auxout{\string
      \newlabel{#1}{{\@currentlabel}{\thepage}}}}}\@gtempa
   \if@nobreak \ifvmode\nobreak\fi\fi\fi\@esphack}
	\gdef\@eqnlabel{#1}}
\def\@eqnlabel{}
\def\@vacuum{}
\def\draftmarginnote#1{\marginpar{\raggedright\scriptsize\tt#1}}
\def\draft{\oddsidemargin -.5truein
	\def\@oddfoot{\sl preliminary draft \hfil
	\rm\thepage\hfil\sl\today\quad\militarytime}
	\let\@evenfoot\@oddfoot	\overfullrule 3pt
	\let\label=\draftlabel
	\let\marginnote=\draftmarginnote
   \def\@eqnnum{(\theequation)\rlap{\kern\marginparsep\tt\@eqnlabel}%
\global\let\@eqnlabel\@vacuum}  }
\def\preprint{\twocolumn\sloppy\flushbottom\parindent 2em
	\leftmargini 2em\leftmarginv .5em\leftmarginvi .5em
	\oddsidemargin -.5in	\evensidemargin -.5in
	\columnsep .4in	\footheight 0pt
	\textwidth 10in	\topmargin  -.4in
	\headheight 12pt \topskip .4in
	\textheight 7.1in \footskip 0pt
	\def\@oddhead{\thepage\hfil\addtocounter{page}{1}\thepage}
	\let\@evenhead\@oddhead	\def\@oddfoot{}	\def\@evenfoot{} }
\def\numberbysection{\@addtoreset{equation}{section}
	\def\theequation{\thesection.\arabic{equation}}}
\def\underline#1{\relax\ifmmode\@@underline#1\else
	$\@@underline{\hbox{#1}}$\relax\fi}
\def\titlepage{\@restonecolfalse\if@twocolumn\@restonecoltrue\onecolumn
     \else \newpage \fi \thispagestyle{empty}\c@page\z@
	\def\thefootnote{\fnsymbol{footnote}} }
\def\endtitlepage{\if@restonecol\twocolumn \else \newpage \fi
	\def\thefootnote{\arabic{footnote}}
	\setcounter{footnote}{0}}  %\c@footnote\z@ }
\catcode`@=12
\relax
\def\figcap{\section*{Figure Captions\markboth
	{FIGURECAPTIONS}{FIGURECAPTIONS}}\list
	{Figure \arabic{enumi}:\hfill}{\settowidth\labelwidth{Figure 999:}
	\leftmargin\labelwidth
	\advance\leftmargin\labelsep\usecounter{enumi}}}
 \relax
\def\tablecap{\section*{Table Captions\markboth
	{TABLECAPTIONS}{TABLECAPTIONS}}\list
	{Table \arabic{enumi}:\hfill}{\settowidth\labelwidth{Table 999:}
	\leftmargin\labelwidth
	\advance\leftmargin\labelsep\usecounter{enumi}}}
 \relax
\def\reflist{\section*{References\markboth
	{REFLIST}{REFLIST}}\list
	{[\arabic{enumi}]\hfill}{\settowidth\labelwidth{[999]}
	\leftmargin\labelwidth
	\advance\leftmargin\labelsep\usecounter{enumi}}}
 \relax
\makeatletter
\newcounter{pubctr}
\def\publist{\@ifnextchar[{\@publist}{\@@publist}}
\def\@publist[#1]{\list
	{[\arabic{pubctr}]\hfill}{\settowidth\labelwidth{[999]}
	\leftmargin\labelwidth
	\advance\leftmargin\labelsep
	\@nmbrlisttrue\def\@listctr{pubctr}
	\setcounter{pubctr}{#1}\addtocounter{pubctr}{-1}}}
\def\@@publist{\list
	{[\arabic{pubctr}]\hfill}{\settowidth\labelwidth{[999]}
	\leftmargin\labelwidth
	\advance\leftmargin\labelsep
	\@nmbrlisttrue\def\@listctr{pubctr}}}
 \relax
\makeatother
\catcode`\@=11
\def\section{\@startsection {section}{1}{0pt}{-3.5ex plus -1ex minus
 -.2ex}{2.3ex plus .2ex}{\raggedright\large\bf}}
\catcode`\@=12
\newskip\humongous \humongous=0pt plus 1000pt minus 1000pt
\def\caja{\mathsurround=0pt}

\newif\ifdtup
\def\panorama{\global\dtuptrue \openup1\jot \caja
	\everycr{\noalign{\ifdtup \global\dtupfalse
	\vskip-\lineskiplimit \vskip\normallineskiplimit
	\else \penalty\interdisplaylinepenalty \fi}}}
\def\eqalignno#1{\panorama \tabskip=\humongous
	\halign to\displaywidth{\hfil$\displaystyle{##}$
	\tabskip=0pt&$\displaystyle{{}##}$\hfil
	\tabskip=\humongous&\llap{$##$}\tabskip=0pt
	\crcr#1\crcr}}
\def\oldreffmt#1{\rlap{[#1]} \hbox to 2\parindent{}}

\def\figfmt#1{\rlap{Figure {#1}} \hbox to 1in{}}

\let\vev\VEV

\def\beq{\begin{equation}}
\def\eeq{\end{equation}}

\def\bea{\begin{eqnarray}}
\def\eea{\end{eqnarray}}
\catcode`\@=11
\def\eqnarray{\stepcounter{equation}\let\@currentlabel=\theequation
\global\@eqnswtrue
\global\@eqcnt\z@\tabskip\@centering\let\\=\@eqncr
\gdef\@@fix{}\def\eqno##1{\gdef\@@fix{##1}}%
$$\halign to \displaywidth\bgroup\@eqnsel\hskip\@centering
  $\displaystyle\tabskip\z@{##}$&\global\@eqcnt\@ne
  \hskip 2\arraycolsep \hfil${##}$\hfil
  &\global\@eqcnt\tw@ \hskip 2\arraycolsep $\displaystyle\tabskip\z@{##}$\hfil
   \tabskip\@centering&\llap{##}\tabskip\z@\cr}

\def\@@eqncr{\let\@tempa\relax
    \ifcase\@eqcnt \def\@tempa{& & &}\or \def\@tempa{& &}
      \else \def\@tempa{&}\fi
     \@tempa \if@eqnsw\@eqnnum\stepcounter{equation}\else\@@fix\gdef\@@fix{}\fi
     \global\@eqnswtrue\global\@eqcnt\z@\cr}

\catcode`\@=12
\font\tenbifull=cmmib10 % bold math italic
\font\tenbimed=cmmib10 scaled 800
\font\tenbismall=cmmib10 scaled 666
\relax

\def\journal{\topmargin .3in	\oddsidemargin .5in
	\headheight 0pt	\headsep 0pt
	\textwidth 5.625in % 1.2 preprint size  %6.5in
	\textheight 8.25in % 1.2 preprint size 9in
	\marginparwidth 1.5in
	\parindent 2em
	\parskip .5ex plus .1ex		\jot = 1.5ex}
\journal

\catcode`\@=11
\def\marginnote#1{}
\newskip\humongous \humongous=0pt plus 1000pt minus 1000pt
\def\caja{\mathsurround=0pt}

\newif\ifdtup
\def\panorama{\global\dtuptrue \openup1\jot \caja
	\everycr{\noalign{\ifdtup \global\dtupfalse
	\vskip-\lineskiplimit \vskip\normallineskiplimit
	\else \penalty\interdisplaylinepenalty \fi}}}
\def\eqalignno#1{\panorama \tabskip=\humongous
	\halign to\displaywidth{\hfil$\displaystyle{##}$
	\tabskip=0pt&$\displaystyle{{}##}$\hfil
	\tabskip=\humongous&\llap{$##$}\tabskip=0pt
	\crcr#1\crcr}}
\font\tenbifull=cmmib10 % bold math italic
\font\tenbimed=cmmib10 scaled 800
\font\tenbismall=cmmib10 scaled 666
\let\vev\VEV

\def\thefootnote{\fnsymbol{footnote}}
\begin{document}
\begin{titlepage}
\begin{center}
\today          \hfill
\hfill    LBL-37893 \\
         \hfill    UCB-PTH-95/35 \\

\vskip .5in

{\large \bf A New Supersymmetric Framework For \\ Fermion Masses.}
%\footnote{This work was supported by the Director, Office of Energy
%Research, Office of High Energy and Nuclear Physics, Division of High
%Energy Physics of the U.S. Department of Energy under Contract
%DE-AC03-76SF00098.}
%alternate footnote for faculty:
\footnote{This work was supported in part by the Director, Office of
Energy Research, Office of High Energy and Nuclear Physics, Division of
High Energy Physics of the U.S. Department of Energy under Contract
DE-AC03-76SF00098 and in part by the National Science Foundation under
grant PHY-90-21139.}

\vskip .5in
N. Arkani-Hamed,
H.-C. Cheng
and
L.J. Hall

{\em Theoretical Physics Group\\
    Lawrence Berkeley National Laboratory\\
and\\
      University of California\\
Department of Physics\\
    Berkeley, California 94720}
\end{center}

\vskip .5in

\begin{abstract}
%insert abstract here
Supersymmetric theories involving a spontaneously broken flavor symmetry can
solve the flavor-changing problem while having quark and lepton masses derived
from both $F$ and $D$ terms. As an example, a theory of leptons is constructed
in which holomorphy constrains the electron to be massless at tree level. The
electron flavor symmetries are broken by D terms, leading to flavor mixing in
the slepton mass matrices, which allows a radiative electron mass to be
generated by the gauge interactions of supersymmetric QED. Such a radiative
origin for the electron mass can be probed by searches for $\tau \rightarrow e
\gamma$, and could be verified or eliminated by measurements of slepton pair
production.

\end{abstract}
\end{titlepage}
%THIS PAGE (PAGE ii) CONTAINS THE LBL DISCLAIMER
%TEXT SHOULD BEGIN ON NEXT PAGE (PAGE 1)

\newpage
\renewcommand{\thepage}{\arabic{page}}
\setcounter{page}{1}
%THIS IS PAGE 1 (INSERT TEXT OF REPORT HERE)
{\bf 1.}
The standard model of particle physics gives no understanding of the pattern of
quark and lepton masses and mixings; for example, why is the electron so light?
The minimal supersymmetric extension of the standard model, while providing
considerable insight into the origin of electroweak symmetry breaking, has made
no progress whatever on this fermion mass problem.
For each measured mass or mixing angle there is a corresponding Yukawa
coupling, which, as in the standard model, must simply be chosen to fit the
data. In this letter we propose an alternative framework for fermion masses
that leads to an effective supersymmetric theory at the weak scale in which the
fermion masses of the lightest generation are not described by Yukawa
couplings; rather they arise as radiative effects when the superpartners are
integrated out of the theory.

The puzzle of the quark and lepton masses can be described
in terms of the pattern of flavor symmetry breaking.
Consider the Yukawa interaction $\lambda_e\bar{e}_Le_Rh$
responsible for the electron mass, which
breaks the independent $U(1)$ phase rotations of $e_L$ and
$e_R$. The smallness of the electron mass requires $\lambda_e$ to be small,
implying that these flavor symmetries are only very weakly broken in nature.
Why? One attractive possibility is that this flavor symmetry breaking, and
therefore the electron mass itself, occurs only as a radiative correction.
In the very paper of 1971 in which spontaneously broken gauge theories were
shown to be renormalizable [1], it was also remarked that certain mass ratios
might be generated by calculable radiative corrections.
Soon afterwards, a theory was constructed in which $m_e/m_\mu$ occurred as an
$O(\alpha)$ radiative effect, and it is instructive to recall two crucial
aspects of this scheme [2].

1. There is a flavor symmetry, $G_f$, which allows only one independent Yukawa
   coupling.
The form of this Yukawa interaction is such that, even when $G_f$ is completely
broken, a mass is generated only for one fermion, identified as the muon, while
the other remains massless at tree level due to an accidental electron flavor
symmetry of this Yukawa interaction.

2. Other interactions of the theory, involving new particles,
break this accidental electron flavor
   symmetry, and appear in loop diagrams to generate $m_e/m_\mu \approx
   O(\alpha)$.

These requirements are easily extended to apply to any case where it is
desired to obtain a fermion mass or mixing angle purely from radiative
corrections.
In the model of Reference 2, $G_f$ was obtained by extending the electroweak
gauge symmetry to $SU(3)_L \times SU(3)_R$, so that electrons and muons
appeared
in the same irreducible multiplet.
When this is broken to the usual $SU(2) \times U(1)$ electroweak symmetry, the
single Yukawa coupling leads to a mass only for the muon.
The electron mass is generated by a loop diagram involving the broken gauge
interactions, with the heavy gauge bosons appearing in the loop.
This model satisfies the above two requirements, allowing
an understanding of $m_e/m_\mu$ as an $O(\alpha)$ radiative effect.

This model also serves to illustrate the difficulties which have plagued
attempts to use radiative corrections to understand the fermion mass spectrum.

A) It is not easy to construct Yukawa interactions which satisfy the first
requirement. In the above model it involves a special vacuum alignment, which
requires a considerable complication of the theory.

B) There is very little motivation for the new flavor symmetry breaking
interactions and exotic particles of the second requirement. In the above model
there is a large extension of the electroweak gauge group which involves doubly
charged gauge bosons and which is not easily extended to the quark sector.

C) The size of the radiative fermion mass cannot be predicted because it
depends on mass ratios of the new exotic particles. Furthermore, these exotic
particles may all be made arbitrarily heavy so that the scheme may not have any
testable consequences.

It is perhaps for these reasons that the idea
of radiative fermion masses has not been as successful as originally hoped.
In this letter we argue
that theories which incorporate weak scale supersymmetry possess features which
allow all three of the above difficulties to be addressed:

A) The Yukawa couplings of the superpotential are not only restricted by $G_f$,
   but also by holomorphy.

B) The accidental flavor symmetries of the Yukawa interactions are typically
   broken by the supersymmetric gauge interactions of $SU(3)\times SU(2) \times
   U(1)$:
there is no need to postulate any new interactions beyond those
required by supersymmetry.
Furthermore, the particles in the loop are just the superpartners of the known
gauge bosons, quarks and leptons.

C) The hierarchy problem dictates that these superpartners are lighter than
about
1 TeV, so that a supersymmetric scheme for radiative masses necessarily leads
to other
testable consequences.

In this letter we outline a general framework for flavor in supersymmetric
theories which follows from imposing a flavor symmetry, $G_f$, and a pattern
for
its breaking. We briefly summarize which quark and lepton masses and mixing
parameters can be obtained
radiatively in this framework, and which must occur at tree level.
We illustrate our ideas with a few simple explicit models for the lepton
sector.
A further discussion of the models, and a complete discussion of the quark
sector, is given in a companion article [3]. We conclude by stressing that a
radiative origin for $m_e$ can be experimentally tested at future accelerators.

{\bf 2.}
At energy scales much larger than the scale of supersymmetry breaking
$\tilde{m}$, the
theory is supersymmetric and the non-renormalization theorems guarantee
that the only corrections to fermion masses occur via wavefunction
renormalizations.
These corrections cannot give mass to a previously massless fermion, and are
not
important for this letter.
At energy scales well beneath $\tilde{m}$, the effective theory is just that of
the
standard model, and radiative corrections to the fermion masses are similarly
uninteresting.
We are therefore interested in radiative corrections at the scale $\tilde{m}$.

In this letter, and in the companion article [3], we assume that the effective
theory at scale $\tilde{m}$ has the minimal gauge group,
$SU(3) \times SU(2) \times U(1)$, and the minimal supersymmetric field content,
three generations of quarks and leptons and two Higgs doublets.
The flavor symmetry group of the pure gauge interactions is  $ U(3)^5=
\prod_a
U(3)_a \; (a = q, u, d, \ell, e)$ as in the standard model,
where $q$ and $\ell$ are the left-handed quark
and lepton doublets, while $u,d$ and $e$ are the right-handed quark and lepton
weak singlets.

The flavor group $U(3)^5$ is broken by eleven
flavor matrices. Three of these are the Yukawa matrices
$\mbox{\boldmath{$\lambda$}}_\alpha (\alpha = u,d,e)$,
familiar from the standard
model, while the remaining eight matrices involve soft supersymmetry breaking
interactions.
These are the three matrix couplings of trilinear scalar interactions
$\mbox{\boldmath{$\xi$}}_\alpha (\alpha = u,d,e)$
, which have the same $U(3)^5$ transformation
properties as the $\mbox{\boldmath{$\lambda$}}_\alpha$,
and the five scalar mass-squared
matrices ${\bf{m}}^2_a$, which transform differently.
For example, while $\mbox{\boldmath{$\lambda$}}_e$
and $\mbox{\boldmath{$\xi$}}_e$
transform as (3,3) under $SU(3)_\ell \times
SU(3)_e$, ${\bf{m}}^2_\ell$ transforms as 1 + 8 under $SU(3)_\ell$ and
${\bf{m}}^2_e$ as 1 + 8 under $SU(3)_e$.

There is considerable freedom
in assignment of $U(3)^5$ breaking to
$\mbox{\boldmath{$\lambda$}}_\alpha, \mbox{\boldmath{$\xi$}}_\alpha$
and ${\bf{m}}^2_a$.
The standard viewpoint assumes that the origin of all $U(3)^5$ breaking, and
therefore of all fermion masses, resides in the
 $\mbox{\boldmath{$\lambda$}}_\alpha$.
The constraints from rare flavor-changing processes are satisfied by taking
$\mbox{\boldmath{$\xi$}}_\alpha$ proportional to
$\mbox{\boldmath{$\lambda$}}_\alpha$, and each
${\bf{m}}^2_a$ proportional to the unit
matrix, so that the soft operators contain no new information about the
breaking of flavor symmetries.
This approach requires very large hierarchies to be built into
$\mbox{\boldmath{$\lambda$}}_\alpha$;
it also misses the opportunity to make use of the
advantages, outlined earlier, that supersymmetry provides for radiative masses

We consider theories with the most general set of couplings consistent with a
flavor symmetry $G_f$. We do not allow $G_f$ to be an $R$ symmetry, ensuring
that  $\mbox{\boldmath{$\lambda$}}_\alpha$ and
$\mbox{\boldmath{$\xi$}}_\alpha$ transform identically
under $G_f$, and hence have the same rank. We require that this rank be less
than three, at least for some $\alpha$. Even if
$\mbox{\boldmath{$\lambda$}}_\alpha$ and
$\mbox{\boldmath{$\xi$}}_\alpha$ have a zero eigenvalue,
the corresponding fermion
can acquire a mass from radiative corrections at scale $\tilde{m}$ [4].
At 1 loop order there is a single relevant diagram, shown in Figure 1 for the
case of the leptons.
Choosing a basis for the $\ell, \tilde{\ell}, e$ and $\tilde{e}$  fields such
that
$\mbox{\boldmath{$\lambda$}}_e, {\bf{m}}^2_\ell$ and
${\bf{m}}^2_e$ are all diagonal, the
radiative contribution to the lepton masses involve ${\bf{V}}_\ell$ and
${\bf{V}}_e$, the $SU(3)_\ell$ and $SU(3)_e$ breaking
flavor mixing matrices induced at the neutral gaugino vertices
by relative rotations of fermions and scalars.
They also involve the scalar trilinear vertices of strength
$\mbox{\boldmath{$\xi$}}_e + \mu \tan
\beta \mbox{\boldmath{$\lambda$}}_e$, which also break axial
lepton number, allowing a connection
between the $\ell$ and $e$ sectors.
Although $\mbox{\boldmath{$\lambda$}}_e$ and
$\mbox{\boldmath{$\xi$}}_e$ have a zero eigenvalue,
the fermion which is massless at tree level can acquire a mass via this diagram
because of the mixings in ${\bf V}_\ell$ and  ${\bf V}_e$. Above the weak
scale, this crucial information about $SU(3)_\ell \times SU(3)_e$ breaking is
encoded in ${\bf m}^2_\ell$ and ${\bf m}^2_e$.
In a similarly defined basis for the quarks, the radiative contributions to the
quark mass matrices involve the flavor mixing matrices at the gluino vertices:
${\bf V}_{q,u,d}$.
It is clearly attractive to speculate that some of the smaller observed
parameters of the flavor sector have their origin in these radiative
corrections.
For example, if the lightest generation masses all come from this effect, one
might expect $m_u \approx m_d \approx (\alpha_s / \alpha) m_e$ [4].
In Reference 4 it was argued that such a radiative origin for $m_d$ implied
that,
for $\tan\beta \approx 1, B^0_d\bar{B}^0_d$ mixing would be maximal.
Since we now know the mixing is not maximal, such a radiative $d$ quark mass
requires large $\tan\beta$.
For large $\tan \beta$ it is well known that there are sizable radiative
contributions to $m_b$ [5] and to other parameters [6], which can affect grand
unified mass relations. We will not be concerned with such corrections in this
paper, rather we are interested in studying which of the small parameters of
the fermion mass sector can be understood as having an origin which is entirely
radiative.

We study theories of flavor in which all dimensionless couplings are of order
unity. However,
as we will show shortly, not all small parameters in the fermion mass sector
can be understood as being purely due to weak-scale loop factors.
It is still necessary
that the theory at the scale $\tilde{m}$ contain some small parameters in
$\mbox{\boldmath{$\lambda$}}_\alpha$,
$\mbox{\boldmath{$\xi$}}_\alpha$ and ${\bf m}^2_a$.
These small parameters are generated by spontaneously breaking a flavor
symmetry, causing mass mixing between light and heavy generations in such a way
that F terms give rise to small entries in
$\mbox{\boldmath{$\lambda$}}_\alpha$ and
$ \mbox{\boldmath{$\xi$}}_\alpha$,
while D terms give small entries in ${\bf m}^2_a$.

%all couplings of order unity and
%provide a mechanism for generating fermion mass hierarchies such as $m_e:
%%m_\mu:
%m_\tau$.
%We know of only two classes of ideas for implementing this: the origin of the
%small dimensionless parameters can either be from loop factors (``1/16
%%$\pi^2$''
%``radiative '') or from ratios of mass scales (``$v_1/v_2$'', ``tree'') which
%should be generated by some dynamical mechanism which avoids fine tuning.
%Figure 1 illustrates the radiative possibility, and Figure 2 the
%Froggatt-Nielsen mechanism for obtaining Yukawa couplings which involve a mass
%hierarchy.
%In this diagram $L$ and $E$ are very heavy vector leptons with $SU(2)$
%%preserving
%mass of scale $M$.
%They have an order unity Yukawa coupling to the Higgs doublet.
%The light leptons $\ell$ and $e$ possess a flavor symmetry spontaneously
%%broken
%by the vev of fields $\phi_\ell$ and $\phi_e$ of size $V$.
%This allows $\ell/L$ and $e/E$ mass mixing of size $V/M$ ($\bar{L}$ marries
%%$ML
%+ V\ell$) so that the light eigenstate acquires a Yukawa coupling of order
%$(V/M)^2$.
%
%Hence supersymmetric theories of flavor can aim to obtain hierarchies in
%$\mbox{\boldmath{$\lambda$}}_\alpha$
%by tree-mixing with very heavy states, or can aim for radiative hierarchies by
%obtaining the flavor breaking parameters in
%$\mbox{\boldmath{$\xi$}}_\alpha$ and ${\bf{m}}^2_a$.
%In this letter, we argue that while the observed hierarchies cannot be
%%entirely
%radiative, there are simple ways of breaking the flavor symmetries $G_f$ such
%that both tree and radiative hierarchies are simultaneously generated.

At some mass scale $M$, much larger than $\tilde{m}$, we
have a full theory of flavor in which all the dimensionless parameters are of
order unity. This theory is based on some flavor symmetry group, $G_f$, which
acts not only on the three light generations but also on vector-like
generations with mass of order $M$. The scale of $M$ is not important; we
assume only that it is less than both the Planck scale and the messenger scale,
where the superpartners first learn about supersymmetry breaking.
We use only renormalizable interactions to construct the full theory at $M$,
non-renormalizable interactions suppressed by the Planck scale do not alter
our results.
Below the scale $M$, the heavy vector generations are integrated
out of the theory to give a $G_f$ invariant effective theory. In addition to
the fields present at scale $\tilde{m}$, this effective theory contains only
gauge singlet flavon fields $\phi$ whose vevs break $G_f$. The scale of these
vevs could be dynamically determined, for example by the evolution of the soft
$m_\phi^2$ parameters to negative values, and hence does not require the
introduction of small parameters.
In the models presented below, this typically
requires the introduction of trilinear superpotential interactions involving
$\phi$, and is not studied in this paper.
These vevs play a crucial role in the mixing of heavy and light generations.

As an example, consider a light lepton, with states $\ell$ and $e$, which is
prevented by $G_f$ invariance from coupling to the Higgs, $h$. Suppose,
however, that
heavy vector leptons, $L$ and $E$ which have the same gauge properties as
$\ell$ and $e$, have mass terms $[M_L \bar{L} L + M_E \bar{E} E]_F$,
and a $G_f$ invariant interaction with the Higgs $[LE h]_F$.
The flavor symmetry which acts on $\ell$ and $e$ is broken by
vevs $\vev{\phi_\ell} = v_\ell$ and $\vev{\phi_e} = v_e$, leading to
mixing of the heavy and light states via the interactions $[\ell \phi_\ell
\bar{L} \; + \; e \phi_e \bar{E}]_F$. This mixing will
induce a Higgs coupling to
the light state which is small, of order $\epsilon_\ell \epsilon_e$, where
$\epsilon_\ell =v_\ell/M_L$ and $\epsilon_e =v_e/M_E$ . This mechanism
for generating small parameters in $\mbox{\boldmath{$\lambda$}}_\alpha$ was
introduced by
Froggatt and Nielsen using Abelian $G_f$ [7], and is illustrated in Figure 2.
In the present work $G_f$ is taken to be non-Abelian: placing
the lightest two generations in a doublet of some non-Abelian $G_f$
allows a solution to the supersymmetric flavor
changing problem. If supersymmetry breaking spurions are inserted at any of the
vertices of Figure 2, corresponding small entries for
$\mbox{\boldmath{$\xi$}}_\alpha$ are also
generated.

A crucial aspect of the above mechanism is the mixing of light and heavy
states, which we could represent as D terms: $[(1/M_L)
L^\dagger \phi_\ell \ell \; + \; (1/M_E) E^\dagger \phi_e e]_D$.
These interactions involve heavy states and cannot
appear in the effective theory beneath $M$. When they are integrated out of the
theory they produce the effective F term: $(1/M_L M_E)[\ell \phi_\ell \; e
\phi_e \;
h]_F$.

We make use of a similar mass mixing effect to generate
small entries in ${\bf m}^2_a$. Suppose that a heavy vector lepton with mass
term $[M \bar{L}L]_F$ mixes via $\phi$ vevs with two different light states,
which, for reasons that will emerge later, we call $\ell_1$ and $\ell_3$:
$[\ell_1 \phi_3 \bar{L} \; + \; \ell_3 \phi_1 \bar{L}]_F$. In this case
$\bar{L}$
acquires a Dirac mass coupling to a linear combination of
$(L,\ell_1, \ell_3)$, leaving
the two orthogonal combinations massless.
There is an important distinction between $\ell_3$ and $\ell_1$. A tree-level
interaction with the Higgs is present for $\ell_3$: $[\ell_3 e_3 h]_F$, but
not for $\ell_1$. This interaction could either be a tree-level Yukawa coupling
of the full theory, or it could be induced in the effective theory by
Froggatt-Nielsen mass mixing. Of the two orthogonal massless combinations of
$(L, \ell_1, \ell_3)$, one involves only $(L, \ell_1)$ and has no tree level
Higgs coupling, we call it $\ell_e$. The other is mainly $\ell_3$ and does have
a Higgs coupling, we call it $\ell_\tau$. The necessity to rotate from the
flavor basis $\ell_1, \ell_3$ to the mass basis  $\ell_e, \ell_\tau$ is shown
diagrammatically in Figure 3, where integrating out $\bar{L}$ yields
off-diagonal kinetic energy D terms in the flavor basis:
$(1/M^2) \; [\ell_1^\dagger \phi_1 \; \phi_3^\dagger \ell_3]_D$.

What is the consequence for the scalar mass matrix ${\bf m}^2$ of performing
this rotation from flavor to mass basis? If ${\bf m}^2$ was initially
proportional to the unit matrix in the $3 \times 3$ space of
$(L, \ell_1, \ell_3)$ the rotation would have no consequence. However, since
$L$ is not unified with $\ell_1$ or $\ell_3$ in an irreducible representation
of $G_f$, $m^2_{LL}$ and  $m^2_{l_1 l_1}$ are unrelated. This is all that is
required to generate an off-diagonal entry in the mass basis:
$m^2_{e \tau} / m^2 \; \approx \; v_1 v_3/M^2$.
Additional comparable contributions to  $m^2_{e \tau}$ arise when the vertices
of Figure 3 are the soft scalar trilinear interactions rather than the
superpotential interactions.

Theories with heavy vector generations have long been used to generate
hierarchical Yukawa couplings by ``heavy-light" mixing induced by $G_f$
breaking [7]. Perhaps the crucial new observation of this work is that in such
theories ``light-light" mixing is also generated when the heavy generations are
integrated out. This leads to flavor breaking in ${\bf m}^2$ rather than in
$\mbox{\boldmath{$\lambda$}}$,
which can then generate fermion masses by weak-scale radiative
corrections.

%However, the orthogonality
%requirement implies that the light eigenstates are mixtures of $\ell_1$ and
%$\ell_3$. This is represented in Figure 3 by diagrammatically integrating out
%$\bar{L}$ to produce the D term: $(1/M^2) \; [\ell_1^\dagger \phi_1 \;
%\phi_3^\dagger \ell_3]_D$ in the low energy effective theory. The ${\bf
%%m}^2_a$
%matrices start out as the most general consistent with $G_f$ invariance. If
%%$L,
%\ell_1$ and $\ell_3$ are in different representations of $G_f$, then ${\bf
%m}^2_a$ is diagonal in this basis with eigenvalues $(m_L^2, m_1^2, m_3^2)$.
%%The
%state mixing represented by the D term  then introduces an off diagonal
%$m^2_{13}$ entry, leading to a non-zero $V_{13}$. The same result occurs even
%if $\ell_1$ and $\ell_3$ are elements of the same representation of $G_f$
%so $m_1^2 = m_3^2$. In this case,
%if L has the same $G_f$ properties as $\ell_1/\ell_3$ then
%${\bf m}^2_a$ can start off with off-diagonal entries so that non trivial
%%$V_{13}$
%is immediate. If L is in a different representation of $G_f$ then, even though
%the initial ${\bf m}^2_a$ are diagonal, A term contributions ...

In a supersymmetric theory of flavor, where all couplings of the full theory
are
of order unity, the large fermion mass can arise directly from Yukawa
couplings,
but the smaller ones must come either from Froggatt-Nielson mass mixing, or
from weak-scale radiative loops.
In a perturbative theory of flavor, the top mass must come from a tree-level
Yukawa coupling, but one could contemplate the $b$ and $\tau$ masses
originating
from mass mixing or from loops.
In this letter we are interested in the case that some of the light fermion
masses occur radiatively, which we will find requires large
$\tan\beta$, and hence it is reasonable
for $m_b$ and $m_\tau$ to arise  from tree-level Yukawa couplings,
with $m_t/m_b$ described dominantly by $\tan\beta$.

An attractive possibility is for the heaviest generation to occur at tree
level, while the lighter two generations both occur radiatively. One way of
attempting this is to have $G_f$ be an $R$ symmetry, allowing the rank of
$\mbox{\boldmath{$\xi$}}$ to be larger than that of
 $\mbox{\boldmath{$\lambda$}}$ [8,9]. For
example, suppose that $G_f$ requires $\lambda_{22}$ to vanish, while allowing a
non-zero $\xi_{22}$, which could appear in the diagram of Figure 1 yielding
second generation masses.
This would require a large value of $\xi_{22}$, and since $\lambda_{22}$
vanishes, the true vacuum has large electric charge breaking vevs
for the scalars of the second generation [10]. Theories of this sort are
excluded unless it is possible to arrange for the universe to evolve to the
desired, very long lived, metastable vacuum.
Hence, if the only non-zero element of
$\mbox{\boldmath{$\lambda$}}$ is $\lambda_{33}$,
we limit the non-zero elements of $\mbox{\boldmath{$\xi$}}$
to $\xi_{33}$, $\xi_{3i}$ and $\xi_{i3}$, where $i=1,2$.

%Are the second generation masses, for $c, s$ and $\mu$, generated by
%tree-level mass mixing or by
%the weak-scale radiative corrections of Figure 1, proportional to
%$V_{L_{23}} \xi_{33} V_{R{32}}$? ($V_L$ is $V_{\ell}$ or $V_q$, while  $V_R$
%is  $V_e$,  $V_u$ or  $V_d$ depending on the charge sector under study.)
%An attractive possibility is for the heaviest generation masses to occur at
%%tree
%level, and all other masses and the Kobayashi-Maskawa mixing matrix to be
%induced by these loop corrections [6].
%We find that this is not possible if the effective theory at the weak scale is
%the supersymmetric theory of minimal field content.
%The second generation mass cannot come from $V_{L_{22}}\xi_{22}V_{R_{22}}$.
%This vacuum stability argument implies that the only non-zero entry in
%$\mbox{\boldmath{$\xi$}}$ is $\xi_{33}$.

It is straightforward to see that the lightest two generation masses cannot be
radiatively generated from $\xi_{33}$ and non-trivial ${\bf m}^2$ matrices.
The second generation masses could come from $V_{L_{23}}^T \xi_{33}V_{R_{32}}$
(although $m_\mu / m_\tau$ is so large that a sufficient muon mass cannot be
generated).
Can the lightest generation mass now arise from $V_{L_{13}}^T
\xi_{33}V_{R_{31}}$?
In theories with significant flavor mixing angles at gaugino vertices,
flavor-changing phenomenology  requires considerable degeneracy amongst scalars
of a given charge of the first two generations.
In the limit that these scalars are exactly degenerate, an $SU(2)$
flavor symmetry is present and ensures that the electron is exactly massless.
Allowing the non-degeneracies to be as large as flavor-changing phenomenology
allows, generates values for $m_{u,d,e}$ which are well below the observed
values. This is shown explicitly in reference 3, where it is also shown that
non-zero values for $\xi_{3i}$ and $\xi_{i3}$ do not change the conclusion that
{\it it is not possible to obtain masses for both light generations
by radiative corrections}.
This means that a supersymmetric theory of flavor, with minimal field content
at
the weak scale, must use the tree-level mass mixing mechanism to
obtain mass for at least one of the light generations.
If the mixing on the left and right-handed fermions for this generation are
described by the parameters $\epsilon_\ell$ and $\epsilon_e$, then the
effective
Yukawa parameter is of order $\epsilon_\ell\epsilon_e$.

What about the origin of the mass of the remaining generation?
For these to occur from tree-level mass mixing effects, there must be further,
very small, flavor symmetry breaking parameters $\epsilon'_\ell$ and
$\epsilon'_e$. The point of this letter is to demonstrate that there is no
need for any such additional
hierarchical parameters; the mass of the remaining generation can be
radiative.
Hence our picture of the hierarchy of the fermion masses of the  three
generations is:
$$
m_3 \ :\ m_2 \ :\ m_1 = 1: \epsilon_\ell\epsilon_e \ : \
{\epsilon_\ell \epsilon_e \over 16\pi^2}.\eqno(1)
$$
It is a very non-trivial aspect of the structure of supersymmetry that the
parameters $\epsilon_{\ell,e}$, which break the flavor symmetries of the second
generation, also appear at radiative order in the first generation masses.
Below we show through explicit models how this arises in the lepton
sector.
In Reference 3 we extend the theory to incorporate quarks, and
show that the up quark mass can easily
occur radiatively, but a radiative down quark
mass is only just consistent with data  on $\overline{B}B$ mixing. We also find
that while $V_{cb}$ and the CP violating phase of the Kobayashi-Maskawa matrix
could arise purely radiatively, it is not possible for $V_{us}$ and $V_{ub}$ to
both be radiative.

{\bf 3.}
Our first model of lepton flavor is based on a flavor group
$G_f= SU(2)_\ell \times SU(2)_e \times U(1)_A$ acting
only on the lightest two generations.
The only small parameters of the theory are those which break this group, and
hence it is the breaking of this group which contains the essence  of lepton
flavor. We later give extensions to $SU(3)_\ell \times SU(3)_e$.
We consider an effective $G_f$ invariant theory of leptons, in which
the leptons have $SU(2)_\ell \times SU(2)_e$ transformation properties
$\ell_3 (1,1), \ell_A (2,1), e_3(1, 1)$ and $e_a(1,2)$.
The Higgs doublet transforms as $h(1,1)$, and there are just two gauge singlet
flavons, $\phi_{\ell A}(2,1)$ and $\phi_{e_a}(1,2)$,
whose vevs $\vev{\phi_\ell} = v_\ell (1,0)$, $\vev{\phi_e} = v_e (1,0)$
describe the breaking of $G_f$.
The $U(1)_A$ charges are $+1$ for $\ell$ and $\phi_\ell$, $-1$ for $e$
and $\phi_e$, and $0$ for $l_3, e_3$ and $h$.
The $\phi_{\ell,e}$ vevs reduce the rank of $G_f$ by 2,
leaving $U(1)_\mu$, muon number, as an exact unbroken symmetry.
In models with a radiative electron mass occurring by 13 mixing, it is
necessary
for the 23 mixing to be very small to avoid a disastrous rate for the rare
decay $\mu \rightarrow e \gamma$. The
origin of $U(1)_A$ will be discussed later.

The most general superpotential of the effective theory below $M$, which is
quadratic in the lepton fields, is
$$
W_{eff} = \lambda \ell_3 e_3 h + {\lambda'\over M^2} (\ell \phi_\ell)(e\phi_e)
 h, \eqno(2)
$$
where $\lambda$ and $\lambda'$ are two dimensionless parameters of order unity.
The absence of any further terms of higher dimensions can be traced to the fact
that the only holomorphic $G_f$ singlets involving $\ell, \phi_\ell, e$ and
$\phi_e$ are
$(\ell \phi_\ell)$ and $(e\phi_e)$.
We have imposed $R$ parity, which forbids interactions such as
$\ell_3(\ell\phi_{\ell}) (e\phi_e)$.

This superpotential has remarkable features. In particular, it yields a tree
level mass hierarchy
$m_\tau : m_\mu : m_e = 1 : \epsilon_\ell \epsilon_e :0$ where $\epsilon_\ell
= v_\ell/M$ and $\epsilon_e = v_e/M$.
Not only is the electron massless at tree level, but the superpotential
possesses an accidental $U(1)_{\ell_1} \times U(1)_{e_1}$ symmetry, thus
satisfying a general requirement for a theory with a radiative electron mass.
It is holomorphy which yields the accidental electron flavor symmetries of the
Yukawa interactions. Without holomorphy, $ (\ell \phi_\ell^\dagger)
(e\phi_e^\dagger) h$ would be allowed, and would give $m_e \approx m_\mu$.
However, these electron chiral symmetries are not exact accidental symmetries
of the entire effective theory, because they are broken by higher order $D$
terms:
$$
{1\over M} \left[ (\ell^\dagger \phi_\ell)\ell_3 +
(e^\dagger \phi_e) e_3\right]_D.\eqno(3)
$$
In general such $D$ terms would be present both as supersymmetric interactions
which
lead to $e/\tau$ wavefunction mixing, and, with the insertion of
supersymmetry breaking spurion fields, as interactions which
induce soft scalar masses mixing $\widetilde{e}$ and $\widetilde{\tau}$.
In either case, the net effect is to generate 13 and 31 entries of
${\bf{V}}_\ell$ and ${\bf{V}}_e$ which are of order $\epsilon_\ell$
and $\epsilon_e$,  respectively.
The loop diagram of Figure 1 generates the radiative electron
mass leading to the hierarchy of (1).
The breaking of axial lepton number originates from
$\lambda_{33}$ and the breaking of the accidental electron
flavor symmetries comes from the 31 entries of ${\bf{V}}_\ell$ and
${\bf{V}}_e$, yielding
$$
m_e = {\alpha\over 4\pi c^2} \left({A + \mu\tan\beta\over m^2}\right) M_1 \;
I\left({M_1^2\over m^2}\right) V_{\ell_{31}}V_{e_{31}} m_\tau\eqno(4)
$$
where the scalar taus have been taken degenerate with mass $m$ and are assumed
to be much lighter than the selectrons, $M_1$
is the bino mass, $c$ is the cosine of the weak
mixing angle, and $I$ is a dimensionless integral with $I(1) = 1/2$.
Taking $\mu = M_1=m$ gives $m_e = 0.5 $ MeV ($A/m + \tan\beta)
 V_{\ell_{31}}V_{e_{31}}$.
Since $V_{\ell_{31}}V_{e_{31}} \approx \epsilon_\ell \epsilon_e
\approx m_\mu/m_\tau$,
the electron mass is large enough only for large
$(A/m + \tan \beta)$.
The $A$ parameter cannot be large enough to dominate this bracket without
leading to
a vacuum instability, hence we derive the prediction that tan $\beta$ is large
in this scheme:
$$
\tan\beta \approx {1\over \epsilon_\ell\epsilon_e},\eqno(5)
$$
in the range of 10 - 50.
The effective theory of lepton flavor defined by equations (2) and (3)
has different origins for all three lepton masses, and leads to the hierarchies
of (1). The $U(1)_A$ symmetry is a necessary component of $G_f$; without it D
terms, like those of (3) but with $\phi_{\ell,e} \rightarrow
\phi_{\ell,e}^\dagger$,
would occur, giving rise to an unacceptable rate for $\mu \rightarrow e
\gamma$. To avoid this, $U(1)_A$ should act as $L_e + L_\mu$ on lepton fields,
and identically on $\phi_\ell$ and $\ell$, and on $\phi_e$ and $e$.

It is very straightforward to write down the full $SU(2)_\ell \times
SU(2)_e\times U(1)_A$
invariant theory which leads to the effective theory of equations (2) and (3).
The interaction of (2) which leads to the muon mass is obtained by integrating
out a
heavy vector lepton $L_3,\overline{L}_3, E_3, \overline{E}_3$ which is singlet
under $SU(2)_\ell \times SU(2)_e$ but has $U(1)_A$ charges of $+2$ and $-2$ for
$L_3$ and $E_3$. The superpotential is
$$
\eqalignno{
W_1 &= \lambda \ell_3 e_3 h + M_{L_{3}}\overline{L}_3L_3 +
M_{E_{3}}\overline{E}_3
      E_3\cr
    &+ \lambda_L L_3E_3h + \lambda_\ell (\ell \phi_\ell)\overline{L}_3 +
       \lambda_e(e\phi_e)\overline{E}_3.&(6)\cr}
$$
The muon mass is generated by the Froggatt-Nielsen mass mixing diagram of
Figure 2.
The $D$ terms of equation (3) are obtained by integrating out a heavy
vector lepton which has $L$ and $\overline{L}$ transform as (2,1) and $E$ and
$\overline{E}$ transform as (1, 2) under $SU(2)_\ell \times SU(2)_e$.
Under $U(1)_A$, $L$ and $E$ transform as $+1$ and $-1$, so the additional
interactions are
$$
\eqalignno{
W_2 &= M_L\overline{L}L + M_E \overline{E} E + \lambda'_\ell
(\overline{L}\ell)S
+ \lambda''_\ell (\overline{L}\phi_\ell) \ell_3\cr
    &+ \lambda'_e (\overline{E} e)S +
\lambda''_e(\overline{E}\phi_e)e_3&(7)\cr}
$$
where $S$ is a singlet and brackets such as $(\overline{L}\ell)$
denote an $SU(2)$ singlet combination of two doublets.
The $D$ terms which induce 13 mixing and lead to the electron mass
are shown in Figure 3 for the $\ell$ sector (with $\phi_3$ identified as $S$).

$W_1$ possesses an accidental flavor symmetry on the electron, because by
holomorphy the
electron only enters in the combinations $(\ell \phi_\ell)$ and $(e\phi_e)$.
This accidental symmetry is broken in $W_2$ by the appearance of both
$(\overline{L}\ell)$
and $(\overline{L}\phi_\ell)$ invariants.
Nevertheless $W_2$ does not lead to a tree level electron mass since it
does not contain the Higgs
field.
Adding higher dimension operators, scaled by powers of $(1/M_{Pl})$, does
not alter this argument.

While the $G_f = SU(2)_\ell \times SU(2)_e \times U(1)_A$ models described
above
provide a very simple explicit model to illustrate
the origin of muon and electron masses, the passage from $SU(2)$ to $SU(3)$
allows a great simplification in the representation
and Yukawa parameter structure, and also sheds light on the origin of
$U(1)_A$ which leads to unbroken muon number.

The representations of the $G_f = SU(2)_\ell \times SU(2)_e \times U(1)_A$
 theory described above strongly suggest an underlying
$SU(3)_\ell \times SU(3)_e$ structure, since they can be grouped together
in complete $SU(3)$ multiplets:
$$
(3, 1): \; \left({\ell}\atop{\ell_3}\right),
\left( {\phi_\ell}\atop{\phi_{\ell_3}}\right),
\left( {\overline{L}}\atop{\overline{L}_3}\right); \; \;
 (\overline{3}, 1): \;
\left( {L}\atop{L_3}\right)
$$
$$
(1, 3): \; \left({ e}\atop{e_3}\right),
\left( {\phi_e}\atop{\phi_{e_3}}\right),
\left({\overline{E}\atop{\overline{E}_3}}\right); \; \;
(1, \overline{3}): \;
\left({E}\atop{E_3}\right)\eqno(8)
$$
with the singlet field $S$ of (7) becoming $\phi_{\ell_3}$ in the $\ell$
 sector and $\phi_{e_3}$ in the $e$ sector.
The ten  interactions of $W_1 + W_2$ which do not involve the Higgs
doublet $h$, can be written
in terms of four $SU(3)_\ell \times SU(3)_e$ invariants:
$$
W_3 = M_L \overline{L}L+M_E\overline{E}E+ \lambda_\ell
(\ell\phi_\ell\overline{L})
+ \lambda_e (e\phi_e\overline{E})\eqno(9)
$$
where all fields are now $SU(3) \; \underline{3}$ or $\underline{3}^*$.
The $\overline{L}L$ mass term gives a degenerate mass to both the $SU(2)$
doublet and singlet heavy lepton.
The $\lambda_{\ell}(\ell\phi_\ell \overline{L})$
interaction, which involves an $SU(3)$ epsilon symbol, incorporates all three
of
the interactions
$\lambda_\ell, \lambda'_\ell$ and $\lambda''_\ell$ occurring in (6) and (7).
The passage from $SU(2)$ to $SU(3)$ flavor symmetries therefore yields a
unification of the
mechanisms for the origin of $m_\mu$ and $m_e$.
The exchange of the $SU(2)$ singlet heavy lepton which generates $m_\mu$ is
partnered by the exchange of the heavy $SU(2)$ doublet
lepton which generates $m^2_{13}$, which leads to $m_e$.

The unifications of the mass mixings with heavy leptons required for $m_\mu$
and $m_e$ generation suggests that a true prediction
for $m_e/m_\mu$ might be possible.
We have been unable to accomplish this because the diagrams for
$m_\mu$ and $m_e$ involve different couplings to the Higgs boson h.
While we find $W_3$ to be a convincing set of interactions to describe the
mixing of heavy and light leptons, it is incomplete for
a theory of flavor with
$G_f = SU(3)_\ell \times SU(3)_e$ for two reasons:

1. There must be further $SU(3)_\ell \times SU(3)_e$ interactions which involve
the Higgs doublet $h$.
These should lead to the interactions $\lambda \ell_3e_3h + \lambda_L L_3E_3h$
of equation (6).

2. The two flavor multiplets, $\phi_\ell (3, 1)$ and $\phi_e(1, 3)$ of equation
(8),
are insufficient to break $SU(3)_\ell \times SU(3)_e$.
We have assumed these fields to have vevs $(v_1, 0, v_3)_{\ell, e}\;$;
but $SU(3)$ rotations could put these into the form $(0,0, v_3)_{\ell, e}$.
It is necessary to introduce further flavons which have vevs
which serve to define the third direction.

There are many ways to satisfy the above, depending on the $SU(3)_\ell \times
SU(3)_e$ transformation properties chosen for the
Higgs doublet and for the additional flavons, and below we give a
straightforward example.
An effective theory which accomplishes points 1 and 2 is obtained by adding
flavons $\Phi (3,3)$ and $\overline{\Phi} (\overline{3}, \overline{3})$
with vevs $\Phi_{33}$ and $\overline{\Phi}_{33}$ being non-zero.
Keeping the Higgs doublet $h$ as a singlet, $(1, 1)$,
two effective interactions can be written
$$
W_4 = {\widetilde{\lambda}\over M} (\ell \overline{\Phi} e) h +
{\widetilde{\lambda}_L\over M} (L\Phi E)h.\eqno(10)
$$
Inserting $\Phi$ and $\overline{\Phi}$ vevs, the $\widetilde{\lambda}$ and
$\widetilde{\lambda}_L$ interactions generate the required $\lambda$
and $\lambda_L$ interactions of (6).

Finally we wish to give the full theory behind (10): what heavy particles of
mass $M$
must be introduced?
The simplest possibility is that there are extra heavy pairs of Higgs doublets.
In addition to (8), the fields of the full theory are:
$$
\eqalignno{
(3, 3): &\Phi, H, \overline{H}' \hskip .25in (\overline{3}, \overline{3}):
    \overline{\Phi}, \overline{H}, H'\cr
(1, 1): &h, \overline{h}.&(11)\cr}
$$
Here $H$ and $H'$ have the same gauge quantum numbers as $h$, and
$\overline{H}$ and $\overline{H}'$ the same as $\overline{h}$.
The interactions beyond $W_3$ are
$$
\eqalignno{
W_5 &= M_H \overline{H} H+ M_{H'} \overline{H}'H' + \lambda_E LEH + \lambda_e
    \ell e H'\cr
    &+ \lambda_H\overline{H}\Phi h + \lambda'_H \overline{H}'
\overline{\Phi}h&(12)\cr}
$$
as well as unimportant couplings involving $\overline{h}$
and trilinear $\overline{H}'\Phi H$ type couplings.
On integrating out the heavy Higgs $H$ and $H'$, the interactions of (12)
generate the effective interactions of (10), as shown in Figure 4.
This illustrates how Froggatt-Nielsen mass mixing can occur in the Higgs
sector.

In our view, the generation of $\ell_3e_3h$ and $LEh$ from (12) is
not as elegant as the lepton mass mixing for $m_e$ and $m_\mu$ induced by (9).
Nevertheless the complete theory with fields (8) + (11) and interactions
$W_3 + W_5$ allows us to address two further questions: the origin of
$U(1)_A$ and of $R$ parity.

We take the full theory to have the most general set of interactions amongst
the
fields of (8) $+$ (11) which are invariant under $G_f = SU(3)_\ell \times
SU(3)_e$.
There are four holomorphic, $G_f$ invariants involving the fields of (8),
as shown in (9).
Any higher dimension operator would just involve products of these.
These interactions possess an accidental $U(1)_\ell \times U(1)_e$ symmetry
where,
under $U(1)_\ell, \ell $ and $\phi_\ell$ have charge $+1$, and $\overline{L}$
and $L$ have charge $-2$ and $+2$.
Similarly, under $U(1)_e, e$ and $\phi_e$ have charge $+1$ and $\overline{E}$
and $E$ have charge $-2$ and $+2$.
(Combining the epsilon symbols with $\overline{L}$ and with $\overline{E}$,
these symmetries can be understood as trialities and are $U(1)$s contained in
$U(3)_\ell \times U(3)_e)$.
When the Higgs multiplets are added, the only interactions of (12) which
break these accidental symmetries are the ones involving the leptons.
At the renormalizable level, these all have the form ``$\ell e h$" involving
one ``$\ell$" and one ``$e$".
Hence, these break $U(1)_\ell \times U(1)_e$ to the axial combination,
which is just lepton number on the lepton fields.
After further breaking, this becomes precisely $U(1)_A$ on the fields of the
effective theory below $M$, leading to muon number conservation.
In the context of $SU(3)$ flavor symmetries, we see that the conservation of
muon number is not an ad hoc constraint placed on the theory, but arises as an
automatic consequence of the simple theory of equation (9).

In the minimal supersymmetric standard model,
and also in our $G_f = SU(2)_\ell \times SU(2)_e \times U(1)_A$ model, $R$
parity must be imposed
by hand.
However, since $R$ parity acts on lepton fields it should surely be understood
from the flavor symmetry.
In the $SU(3)_\ell \times SU(3)_e$ model of $W_3+W_5$, $R$ parity is an
accidental
symmetry. In the minimal supersymmetric theories, and also in the $G_f =
SU(2)_\ell
 \times SU(2)_e \times U(1)_A$ theory, there is no
symmetry distinction between $\ell$ and $h$: $R$ parity must be imposed by hand
to provide an artificial distinction to avoid too much lepton number violation.
However, in the $G_f = SU(3)_\ell \times SU(3)_e$ model, there is a distinction
built
into the $G_f$ structure, with $\ell$ transforming as (3, 1) and $h$ as (1, 1).
Even allowing for mass mixing; the heavy leptons $L$ transform as
$(\overline{3}, 1)$
and the heavy Higgs $H$ as $(3, 3)$, maintaining sufficient difference between
lepton
and Higgs sectors that $R$ parity violating couplings are all forbidden by
$G_f$
at the renormalizable level.

We have taken the theory at $M$ to be renormalizable, however, $G_f$ invariant
non-renormalizable operators scaled by powers of $M_{Pl}$ are to be expected.
In the $SU(3)_l \times SU(3)_e$ model, these will lead to small violations of
muon number and $R$ parity, suppressed by powers of $(M/M_{Pl})$.
Alternatively, $U(1)_A$ could be promoted to an exact symmetry by extending
$G_f$ to $U(3)_\ell \times U(3)_e$.

{\bf 4.}
In this letter we have proposed a new framework for understanding flavor in
supersymmetric theories which have a flavor symmetry $G_f$ spontaneously broken
by a set of vevs $\vev{\phi}$. When heavy vector generations of mass $M$ are
integrated out of the theory, $G_f$ breaking interactions are generated which
depend on the set of small parameters $\epsilon = \vev{\phi}/M$: scalar masses
from
D terms and Yukawa couplings from F terms. We have constructed theories of the
lepton sector where these Yukawa couplings lead to a muon mass of order
$\epsilon^2 \; m_\tau$, but, because the F terms are holomorphic, the electron
remains massless at tree level. The $G_f$ breaking scalar masses lead, via the
diagram of Figure 1, to a radiative electron mass of order $\epsilon^2 \;m_\tau
/16 \pi^2$.

We find such a theory of lepton flavor, for example the one defined by
equations (2) and (3), to be simple and plausible. If superpartners are
discovered, the proposed radiative mechanism for the electron mass can be
tested quantitatively. The flavor-violating neutralino mixing matrix entries
$V_{\ell_{31}}$ and $V_{e_{31}}$, together with the superpartner spectrum,
could
be measured in the reactions $e^+ e^- \rightarrow \tilde{\tau}^+\tilde{\tau}^-,
\tilde{\tau}^+\tilde{e}^-, \tilde{e}^+\tilde{\tau}^-,
\tilde{e}^+\tilde{e}^-$, as will be demonstrated elsewhere. Furthermore, the
framework predicts large $\tan \beta$ and an observable decay rate for $\tau
\rightarrow e \gamma$ [3].

\noindent{\bf{References}}
\begin{enumerate}
\item G. t'Hooft, {\it Nucl. Phys.} {\bf B35} 176 (1971).
\item S. Weinberg, {\it Phys. Rev. } {\bf D5} 1962 (1972); H. Georgi and S.
Glashow, {\it Phys. Rev.} {\bf D7} 2457 (1973).
\item N. Arkani-Hamed, H.-C. Cheng and L.J. Hall, LBNL-37894 (1995).
\item L. J. Hall, A. Kostelecky and S. Raby, {\it Nucl. Phys.} {\bf B267} 415
(1986).
\item L. J. Hall, R. Rattazzi and U. Sarid, {\it Phys. Rev.} {\bf D50} 7048
(1994).
\item T. Blazek, S. Raby and S. Pokorski, OHSTPY-HEP-T-95-007 (1995).
\item C. D. Froggatt and H. B. Nielsen, {\it Nucl. Phys.} {\bf B147} 277
(1979).
\item A. Lahanas and D. Wyler, {\it Phys. Lett} {\bf 122B} 258 (1983).
\item T. Banks, {\it Nucl.Phys.} {\bf B303} 172 (1988).
\item J. Fr\`{e}re, D. R. T. Jones and S. Raby, {\it Nucl.Phys.} {\bf B222}
11 (1983).
\end{enumerate}

\noindent{\bf Figure Captions}

1. A radiative diagram for lepton masses involving internal superpartners.

2. Mass mixing from flavon vevs, $\vev{\phi_{l,e}}$, induces a Higgs coupling
   to a light lepton.

3. Mass mixing from flavon vevs, $\vev{\phi_{1,3}}$, induces a flavor changing
   $D$ term for the light leptons.

4. Mass mixing from flavon vev, $\vev{\bar{\Phi}}$, induces a tau Yukawa
   coupling in the $SU(3)_\ell \times SU(3)_e$ model.
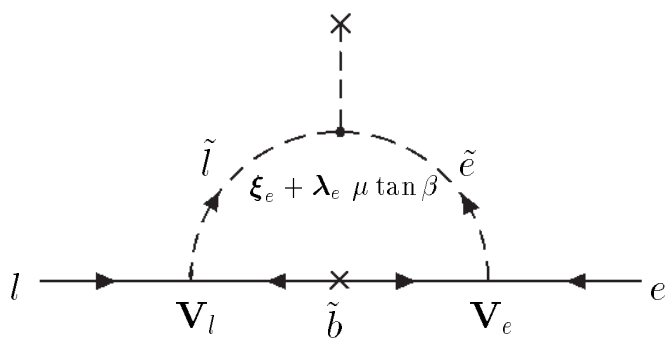
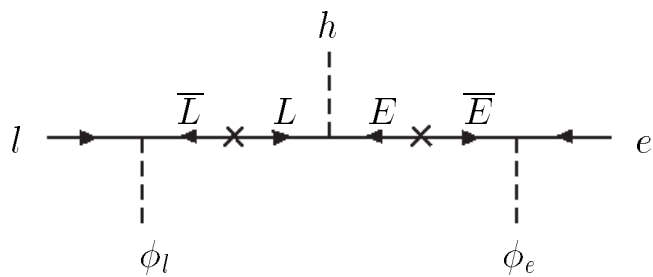
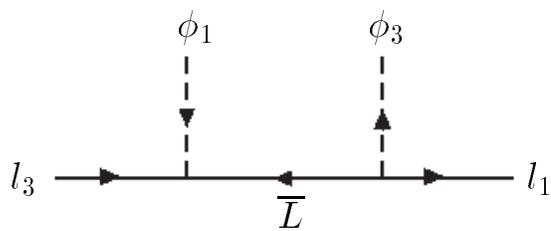
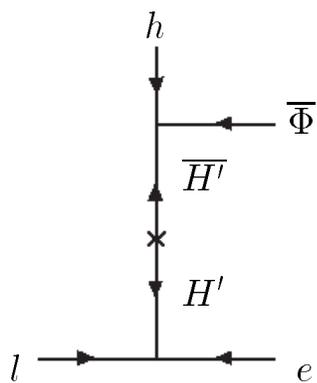
\end{document}